# Chaotic spin precession in antiferromagnetic domain walls

Yangyi Chen[1, 2], Xu Ge[1#], Wei Luo[1], Shiheng Liang[3], Xiaofei Yang[1], Long You[1], Yue Zhang[1*]

1. School of Optical and Electronic Information, Huazhong University of Science and Technology, Wuhan, China

2. State Key Laboratory of Precision Measurement Technology and Instruments, Department of Precision Instrument, Tsinghua University, Beijing, China

3. Department of Physics, Hubei University, Wuhan, China

*Corresponding author: yue-zhang@hust.edu.cn (Yue Zhang)

# Xu Ge has the same contribution to Yangyi Chen.

**Abstract**

In contrast with rich investigations about the translation of an antiferromagnetic (AFM) texture, spin precession in an AFM texture is seldom concerned for lacking an effective driving method. In this work, however, we show that under an alternating spin-polarized current with spin along the AFM anisotropy axis, spin precession can be excited in an AFM DW. Especially, chaotic spin precession occurs at moderate interfacial Dzyaloshinskii-Moriya interaction (DMI), which contributes to a nonlinear term in the dynamic equation of DW precession. Also, crisis-induced intermittent chaos appears when the current density is higher than a critical value. This work not only paves a way to unravel rich spin precession behaviors in an AFM texture but also provides guidelines for developing ultrafast spintronic devices based on new physical principles.

Chaos is deterministic nonlinear dynamics that is sensitive to variation of initial conditions, and it attracts enduring research interests owing to its wide existence in nature and valuable applications, such as weather forecast, secure communication, and neuromorphic computing [1 ~ 3].

Nonlinear terms in dynamic equations are necessary for chaos. In a magnetic medium, spin dynamics is governed by Landau-Lifshitz-Gilbert (LLG) equation, which includes inherent nonlinear terms from dipole-dipole interaction and magnetic anisotropy. Therefore, chaotic spin dynamics is usually observed in nanoscale spintronic devices, such as current-induced chaotic precession of spatially uniform magnetization in a spin valve [4 ~ 8].

In addition to uniform magnetization, magnetic textures with inhomogeneous magnetization, such as domain walls (DWs), skyrmions, vortexes, and bimerons, also widely exist in magnetic materials. The spin dynamics of a magnetic texture generally includes translation (displacement of texture center) and precession (spin precession in a texture) [9 ~ 12]. In an antiferromagnetic (AFM) medium, most attention is paid to texture translation instead of precession up until now. A typical example is ultrafast AFM DW translation (velocity > 1000 m/s) with an upper limit of velocity, which is analogous to the relativistic motion of a massive object [13 ~ 17]. Very recently, chaotic translation was also predicted in an AFM texture under an alternating spin current by introducing a nonlinear force originated from the boundary of an AFM medium [18 ~ 22].

In an AFM material, DW precession is absent for uniform DW translation [13, 23, 24]. This ensures stable DW translation at a high velocity without worrying about Walker breakdown for an FM DW [25]. On the other hand, however, strong exchange coupling in an AFM DW offers potential for ultrafast spin precession in the DW, and this DW precession can exhibit special spin dynamics behaviors owing to the unique inertia of an AFM DW [25]. Nevertheless, how to excite AFM DW precession is still an open question.

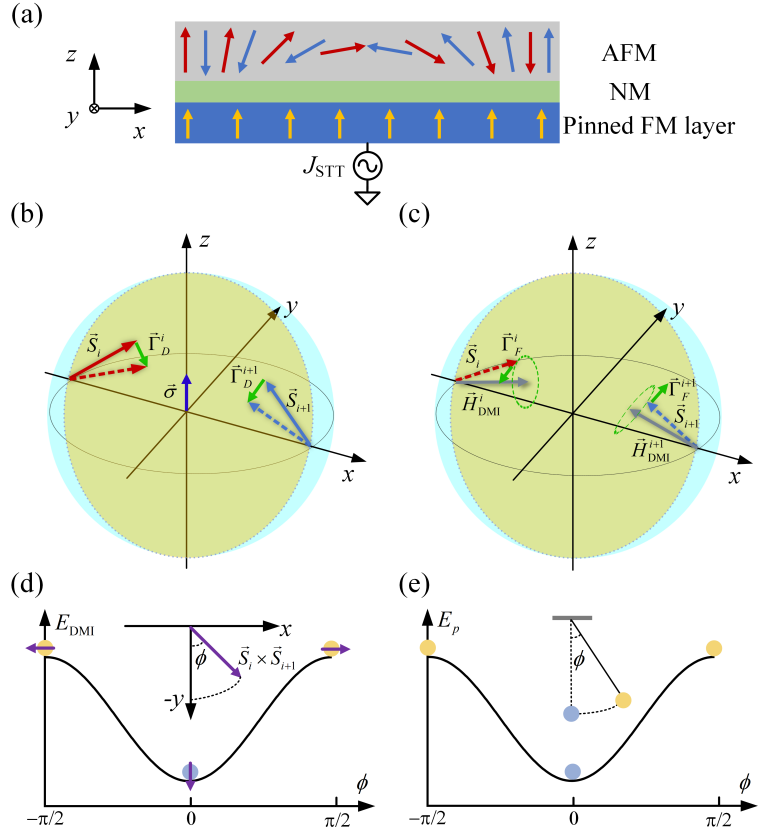

**Figure 1.** (a) Schematic of the model for exciting precession of an AFM DW: Injection of an alternating spin-polarized current into an AFM layer with a Néel-type DW structure. Here the spin-polarized current with spin along *z*-axis is generated by the injection of an alternating current passing through the pinned FM layer; (b) rotation of two spin moments ($\vec{S}_i$ and $\vec{S}_{i+1}$) near the center of an AFM DW under damping-like spin transfer torque (STT); (c) precession of $\vec{S}_i$ and $\vec{S}_{i+1}$ under field-like torque of DMI field; [The yellow regions in (b) and (c) indicate the *xz* plane.] This precession gives rise to the precession of $\vec{S}_i \times \vec{S}_{i+1}$ in the *xy* plane with the variation of energy in terms of cos$\phi$ with $\phi$ the angle between $\vec{S}_i \times \vec{S}_{i+1}$ and $D\vec{e}_y$ (d). This is analogous to the gravitational potential energy of a pendulum (e).

In this work, we predict AFM DW precession excited by an alternating spin-polarized current with spin along the AFM easy anisotropy axis (z direction) (Fig. 1). Especially, the DW precession exhibits chaotic behaviors at moderate interfacial Dzyaloshinskii-Moriya interaction (DMI) that contributes to a nonlinear term in the dynamics equation for DW precession. This work unravels rich behaviors of spin precession in an AFM DW and paves a way for developing nanoscale spintronic devices with a fast processing speed.

We considered a spin-valve-like device composed of a free AFM layer and a pinned FM layer separated by a nonmagnetic metallic layer [Fig. 1(a)]. The FM magnetization was fixed along the *z*-axis direction, and a Néel-type chiral DW was generated in the AFM layer with interfacial DMI.

This device configuration is experimentally feasible since similar devices have been proposed and fabricated to investigate spin dynamics under a spin-polarized current [26, 27]. When an electrical current passes through the pinned FM layer, a spin-polarized current with spin $\vec{\sigma} = \sigma \vec{e}_z$ is generated and injected into the AFM DW. Under a damping-like spin transfer torque (STT) [$\vec{\tau}_{DL} \propto \vec{S} \times (\vec{S} \times J\vec{\sigma})$] acting on spin $\vec{S}$ at current density $J$, the two spins near the DW center rotate downwards or upwards in the $xz$ plane [the yellow planes in Figs. 1(b) and (c)], depending on the sign of $J$ [Fig. 1 (b)]. This $\vec{\tau}_{DL}$ breaks the initial DW stability, and the DW spins can be triggered to precess under the field-like torques from exchange coupling, magnetic anisotropy, DMI, and spin-polarized current. We will verify that the field-like torque of DMI plays a critical role in exciting chaotic DW precession. (S1 in the Supplementary Materials).

The interfacial DMI energy is quantified as $E_{DMI} = D\vec{e}_y \cdot (\vec{S}_i \times \vec{S}_{i+1})$ [28, 29], which gives rise to the effective DMI field acting on $\vec{S}_i$ as $\vec{H}_{DM}^i = -\frac{D_{sim}}{\mu_0 \mu_s} \vec{e}_y \times (\vec{S}_{i+1} - \vec{S}_{i-1})$ with the magnetic permeability in vacuum $\mu_0$ and the magnetic moment $\mu_s$. Here $\vec{S}_i$ is the normalized spin moment at the $i^{th}$ site. The mirror symmetry between $\vec{H}_{DM}^i$ and $\vec{H}_{DM}^{i+1}$ gives rise to opposite chirality for the precession of $\vec{S}_i$ and $\vec{S}_{i+1}$ [Fig. 1(c)]. As a result, $\vec{S}_i \times \vec{S}_{i+1}$ rotates in the $xy$ plane, leading to the variation of DMI energy in terms of $\cos\phi$ with $\phi$ the angle between $\vec{S}_i \times \vec{S}_{i+1}$ and $D\vec{e}_y$ [Fig. 1(d)]. (The angle variation between $\vec{S}_i$ and $\vec{S}_{i+1}$ was disregarded under strong exchange coupling.) This DMI energy is analogous to the gravitational potential energy of a pendulum [Fig. 1(e)]. Since a pendulum can exhibit chaotic mechanical dynamics under an alternating force, we expect chaotic spin precession in an AFM DW under an alternating spin-polarized current.

To confirm our expectation, we derived the dynamic equation of $\Phi$ under an alternating spin-polarized current based on the following procedure [More details about the derivation are in the Supplementary Materials (S1)]. We started with the Hamiltonian $H$:

$$H = A_{sim} \sum_i \vec{S}^{(i)} \cdot \vec{S}^{(i+1)} - K_{sim} \sum_i (\vec{S}^{(i)} \cdot \vec{e}_z)^2 - D_{sim} \sum_i \vec{e}_y \cdot (\vec{S}^{(i)} \times \vec{S}^{(i+1)}). \quad (1)$$

Here $A_{sim}$, $K_{sim}$, and $D_{sim}$ are the exchange energy, anisotropy energy, and DMI energy for every spin. The total magnetization $\vec{m}_i = M_S(\vec{S}_i + \vec{S}_{i+1})$ and Néel vector $\vec{n}_i = \frac{\vec{S}_{i+1} - \vec{S}_i}{2}$ were introduced as the basic quantities for the calculation of AFM dynamics. Here $M_S$ is the saturation magnetization of an AFM sublattice.

In continuous approximation, Eq. (1) was converted into the free energy as:

$$H = S_\perp \int [\frac{a}{2}m^2 + \frac{A}{2}n'^2 + L\vec{m} \cdot \vec{n}' - \frac{K_z}{2}n_z^2 + \frac{D}{2}\vec{e}_y \cdot (\vec{n} \times \vec{n}')]dx. \quad (2)$$

Here the upper prime denotes the derivative over $x$, and $S_\perp$ is the area of the cross-section of an AFM chain. $a$, $A$, $K_z$, and $D$ are homogeneous exchange constant, exchange stiffness constant, anisotropy constant, and DMI constant, respectively. They are related to $A_{sim}$, $K_{sim}$, and $D_{sim}$ by $a = 2A_{sim}/(V_{cell}M_s^2)$; $A = A_{sim}\Delta_x/(wt_z)$; $L = A_{sim}\Delta_x/(V_{cell}M_s)$; $D = 2D_{sim}/(wt_z)$, and $K_z = 4K_{sim}/V_{cell}$, where $d$ is the distance between the nearest neighboring spins. $V_{cell} = wt_z\Delta_x$ with $\Delta_x = 2d$ for the cell size in the $x$ direction, and $w = d$ is the chain width. We considered the parameters of AFM $L_{10}$-MnPt: $d = 0.28$ nm, $A_{sim} = 40$ meV, and $K_{sim} = 0.23$ meV [30 ~ 32]. Here $d$ was calculated based on the lattice structure and lattice constant of MnPt ($a = 0.4$ nm, and $d = a/\sqrt{2}$) [30]. Since the sign and magnitude of interfacial DMI depend on the composition above and below a magnetic layer, the interfacial DMI energy for every magnetic atom can be manipulated in a wide range (from 0 to several meV) [33]. In this work, we show that a DMI energy of $10^{-2}$ meV is enough for exciting chaotic DW precession under the above parameters.

Under the spin-polarized current as indicated in Fig. 1(a), the dynamic equations of $\vec{m}$ and $\vec{n}$ were deduced as [13]:

$$\dot{\vec{n}} = (\gamma \vec{f}_m - G_1 \dot{\vec{m}}) \times \vec{n} + \frac{\gamma B_D}{l} \vec{n} \times (\vec{m} \times \vec{e}_z).  \quad (3)$$

$$\dot{\vec{m}} = (\gamma \vec{f}_n - G_2 \dot{\vec{n}}) \times \vec{n} + \gamma B_D l\, \vec{n} \times (\vec{n} \times \vec{e}_z).  \quad (4)$$

Here the upper dot indicates the derivative over time, $\gamma$ is the gyromagnetic ratio of electrons [$1.76 \times 10^{11}$ rad/(s·T)], and $l$ is the magnitude of the Néel vector, which is close to $2M_S$ under strong exchange coupling. $G_1$ and $G_2$ are the effective damping coefficients for $\vec{n}$ and $\vec{m}$ as $G_1 = \alpha/l$ and $G_2 = \alpha l$, where $\alpha$ is the Gilbert damping coefficient. $B_D = \mu_B PJ/\gamma e M_S t_z$ is the strength of the damping-like STT, where $\mu_B$ is the Bohr magneton; $P = 0.48$ is the spin polarization for Pt/Co [34]; $e$ is the electron charge; $M_S = 5.05 \times 10^5$ A/m, which was calculated based on the magnetic moment of the Mn atom and the volume of space that every Mn occupies in $L_{10}$-MnPt [31]. $t_z = 0.4$ nm is the thickness of the AFM layer. $\vec{f}_m$ and $\vec{f}_n$ are the effective field for $\vec{m}$ and $\vec{n}$:

$$\vec{f}_m = -\delta_m H = -a\vec{m} - L\frac{\partial \vec{m}}{\partial x} \quad \text{and} \quad \vec{f}_n = -\delta_n H = A\vec{n}'' + L\vec{m}' + K_z n_z \vec{e}_z + D\vec{e}_y \times \vec{n}'.$$

After a series of algebra calculations (S1 in the Supplementary Materials), we derived the dynamic equation of $\vec{n}$ as:

$$\ddot{\vec{n}} + a\gamma G_2 \dot{\vec{n}} - a\gamma^2 \left[ A^* \vec{n}'' + K_z n_z \vec{e}_z + D(\vec{e}_y \times \vec{n}') \right] + a\gamma^2 B_D l(\vec{n} \times \vec{e}_z) = \vec{0}. \quad (5)$$

Here $A^* = A - \frac{L^2}{a}$ is the effective exchange constant including the parity-breaking term [35]. In a spherical coordinate system, $\vec{n} = (\sin\theta\cos\Phi,\ \sin\theta\sin\Phi,\ \cos\theta)$, where $\theta$ and $\Phi$ are the polar angle and the azimuthal angle, respectively. We considered a Walker-type DW profile $\theta = 2\arctan[\exp(\frac{x - X(t)}{\lambda})]$ and $\Phi = \Phi(t)$, where $\lambda$ is the DW width. For simplicity, we neglected the variation of $\lambda$ and assumed that the DW structure is kept in the DW precession (We verified this assumption by atomistic simulation as shown in S2 of the Supplementary Materials). Using the scalar product in Eq. (5) with $\frac{\partial \vec{n}}{\partial X}$ and $\frac{\partial \vec{n}}{\partial \Phi}$ and then integrating over the space of chain, the Thiele equations for X(t) and Φ(t) were finally deduced as:

$$\ddot{X} + a\gamma G_2 \dot{X} = 0 .\qquad(6)$$

$$\ddot{\Phi} + a\gamma G_2 \dot{\Phi} - \frac{\pi}{4\lambda}a\gamma^2 D\sin(\Phi) = a\gamma^2 B_D l .\qquad(7)$$

In the previous calculation of the dynamics of an AFM texture, the driving force was generally acted on X instead of Φ, and Φ was usually assumed to be constant without a nonzero initial angular velocity [13]. Therefore, most attention was paid to DW translation instead of DW precession. However, under a spin-polarized current with spin along the AFM anisotropy axis, the STT term appeared in the dynamic equation of Φ instead of X [Eqs. (6) and (7)]. It is noticed that Eq. (7) is analogous to the dynamic equation of a classical pendulum that can exhibit a chaotic oscillation under an alternating driving force. Therefore, a chaotic DW precession is also possible under an alternating spin-polarized current. To verify our prediction, we numerically solved Eq. (7) using the 4$^{th}$ Runge-Kutta method with a time step of 0.01 ps. Also, we exploited atomistic simulation to study the precession of an AFM DW (S2 in the Supplementary Materials).

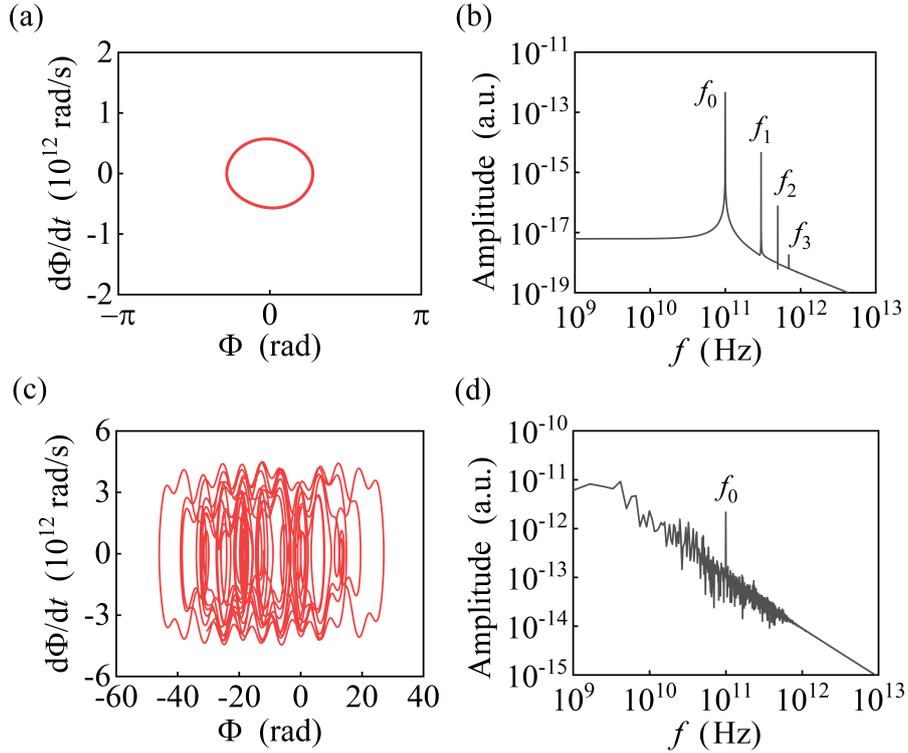

**Figure 2. Trajectories in phase portraits dΦ/dt ~ Φ and spectrum of Φ at $D = -0.02$ meV and $\alpha = 2 \times 10^{-3}$ [(a) and (b)] and at $D = -0.02$ meV and $\alpha = 2 \times 10^{-4}$ [(c) and (d)]. The $f_0$ denotes the frequency of the alternating spin-polarized current. $f_1 = 3f_0$, $f_2 = 5f_0$, and $f_3 = 7f_0$ indicate higher-order DW precessions due to the nonlinear DMI term.**

Representative numerical solutions of Eq. (7) were shown in Fig. 2. Under weak DMI ($D = -0.02$ meV) with $\alpha = 2 \times 10^{-3}$, in addition to synchronized DW precession at frequency $f_0$, weak DW precession at frequencies $f_1 = 3f_0$, $f_2 = 5f_0$, and $f_3 = 7f_0$ was also detected [Figs. 2(a), (b)]. These higher-order signals originated from nonlinear terms in the Taylor expansion of $D\sin\Phi$ in Eq. (7). When $\alpha$ was $2 \times 10^{-4}$, the period of $\Phi(t)$ was absent, and no sharp peak can be seen in the spectrum of $\Phi$. On the other hand, the trajectory in the phase diagram also failed to close after the precession for a long time (longer than 800 ps) [Figs. 2(c) and (d)], which hinted possible chaotic DW precession.

To confirm the transition between periodic and chaotic DW precession with DMI, we calculated the Lyapunov exponents (LEs) [Fig. 3; Details of the LEs calculation are in the Supplementary Materials (S3)] defined as [19]:

$$\text{LE}_i = \lim_{t \to \infty} \frac{1}{t} \ln \frac{\left\| \delta\Phi_t^i \right\|}{\left\| \delta\Phi_0^i \right\|}. \tag{8}$$

Here $\left\| \delta\Phi_0^i \right\|$ is the initial distance between two closed trajectories and $\left\| \delta\Phi_t^i \right\|$ is the distance between two trajectories at time $t$. $i = 1, 2$, and 3 for $\Phi$, $d\Phi/dt$, and $t$, respectively. A positive $\text{LE}_1$ indicates chaotic dynamics with an exponential increase of the distance between two trajectories for any nonzero $\left\| \delta\Phi_0^i \right\|$.

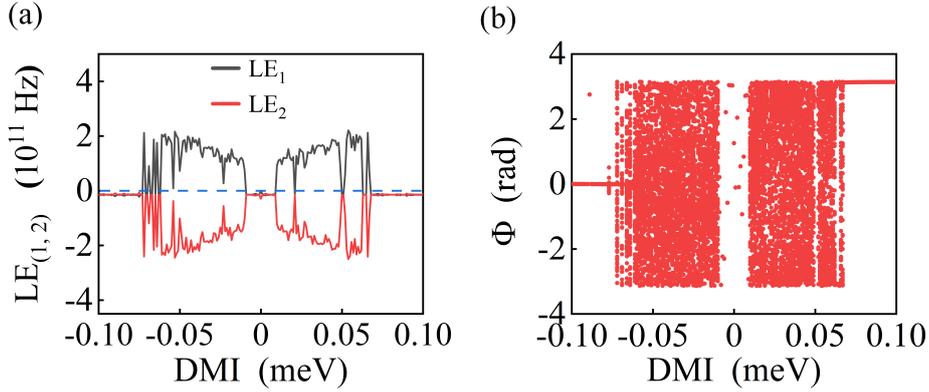

**Figure 3.** (a) $\text{LE}_1$ and $\text{LE}_2$ components of the Lyapunov exponents and (b) bifurcation diagram as a function of DMI energy.

As indicated in Fig. 3(a), $\text{LE}_1$ was negative when the DMI energy was either small (< 0.02 meV) or large (> 0.17 meV). In between, the $\text{LE}_1$ was positive. The $\text{LE}_2$ was symmetric to $\text{LE}_1$, and $\text{LE}_3$ kept zero. This indicates chaotic DW precession under moderate DMI. Under very weak DMI, the nonlinear term in Eq. (7) was negligible. Nevertheless, the DW precession was also depressed under strong DMI. The chaotic DW precession can also be verified from the bifurcation diagram [Fig. 3(b)]. Here we collected $\Phi$ at a series of moments with an interval of the period of the driving current. Only one $\Phi$ appears under either weak or strong DMI, but the number of $\Phi$ approached infinite under moderate DMI. In addition to that, a small tip appeared when DMI was zero. This is ascribed to the irrelevance between $\Phi(t)$ and $\Phi$ in absence of DMI.

In addition to the chaos generated through the bifurcation by manipulating DMI, intermittent chaotic DW precession was also triggered by adjusting the current density. Here, the intermittent chaos depicts alternating periodic and aperiodic dynamics, which can be induced by the collision of a chaotic attractor with the boundary of the basin of its attractor (crisis) in a system with multiple attractors [36, 37]. Here the basin describes the collection of initial conditions that leads to a long-term behavior approaching an attractor. Under typical crisis-induced intermittent chaos, the trajectory in the phase portrait would wander between different attractors.

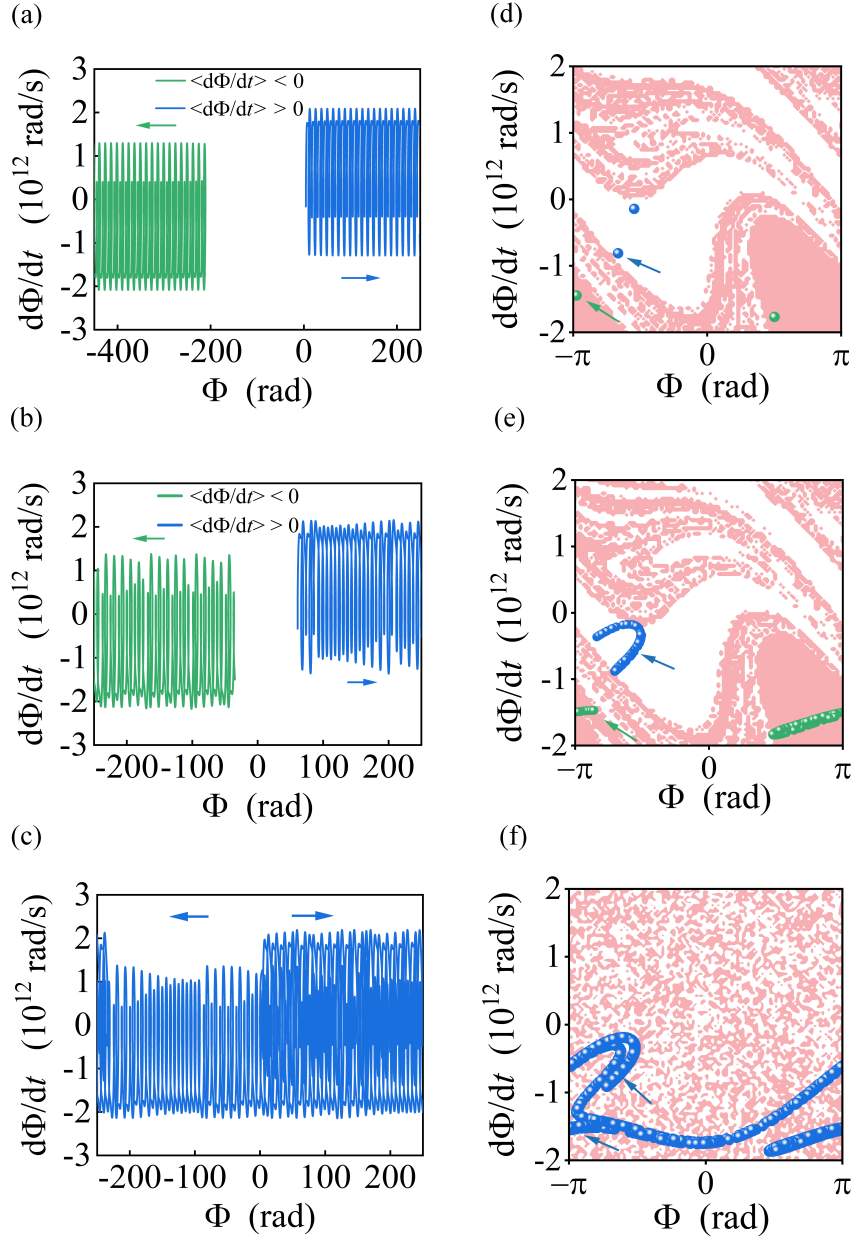

**Figure 4.** Trajectories of phase portraits and the basins of attractors under $J = 4.9 \times 10^{10}$ A/m$^2$ [(a) and (d)], $5.04 \times 10^{10}$ A/m$^2$ [(b) and (e)], and $5.1 \times 10^{10}$ A/m$^2$ [(c) and (f)]. The Poincaré sections at $\omega_0 t = 0$ (mod $2\pi$) under different $J$ were exhibited in (d) ~ (f). The green and blue trajectories in (a) and (c) denote DW precessions with opposite chirality: the $\langle d\Phi/dt \rangle > 0$ (blue) and the $\langle d\Phi/dt \rangle < 0$ (green). These two running modes were separated in (a) ~ (d) but mixed in (e) and (f).

In this work, when $J$ exceeded a critical value $J_{c1} = 4 \times 10^{10}$ A/m$^2$, there were two running attractors with an average positive or negative angular velocity [the blue and green lines in Figs. 4(a) and (b)] as denoted by the white and pink regions in Figs. 4(d) ~ (e), respectively. We collected the temporal $\Phi$ and the basins of attractors under $J > J_{c1}$. When $J = 4.9 \times 10^{10}$ A/m$^2$, the attractors were period-2, and the Poincaré sections at $\omega_0 t = 0$ (mod $2\pi$) were individual dots located within different basins [Figs. 4 (d)]. When $J$ was higher than another critical value $J_{c2} = 4.98 \times 10^{10}$ A/m$^2$, the attractors became chaotic, and the Poincaré sections became continuous lines but were still confined in their own attractors [Figs 4(b) and (e)]. When $J$ was further increased to exceed a third critical value $J_{c3} = 5.04 \times 10^{10}$ A/m$^2$, the two attractors simultaneously collided at the basin boundaries and joined to form a folded chaotic attractor [37], which indicates typical crisis-induced intermittent chaotic DW precession. This can also be verified by the irregular up and down running of $\Phi$ [Figs. 4(c) and (f)].

In summary, we theoretically predict the excitation of chaotic AFM DW precession triggered by a spin-polarized current with spin along the easy anisotropy axis. This chaos originates from the intrinsic DMI of an AFM medium and exhibits particular crisis-induced intermittency. This work unravels rich precession behaviors in an AFM DW and paves a way to develop spintronic devices based on the spin dynamics of an AFM texture.

**Acknowledgment**
The authors acknowledge financial support from the National Key Research and Development Program of China (Grant No. 2022YFE0103300) and the National Natural Science Foundation of China (No. 51971098).